\documentclass[12pt]{article}
\usepackage{times}
\usepackage{geometry}
\usepackage[utf8]{inputenc}
\usepackage{enumitem,amssymb}
\usepackage{ragged2e}
\newlist{thematic}{itemize}{8}
\setlist[thematic]{label=$\square$}
\usepackage{pifont}
\usepackage{amsmath,amsfonts,amssymb}
\usepackage{graphicx}
\usepackage{setspace}
\usepackage{tocloft}
\usepackage{endnotes}

\usepackage{floatrow}

\usepackage{multirow}
\usepackage{array, booktabs}
\usepackage{hyperref}
\usepackage{siunitx}
\usepackage{xcolor}
\usepackage{mnotes}

\usepackage[
backend=biber,
style=ieee,
sorting=ynt,
language=american
]{biblatex}

\newcommand{\xmark}{\ding{55}}%

\newcommand{\flagendnote}[1]{}

\DeclareSIUnit{\mas}{mas}
\DeclareSIUnit\parsec{pc}
\DeclareSIUnit{\arcsec}{asec}
\DeclareSIUnit{\arcmin}{amin}
\DeclareSIUnit\lightyear{ly}
\DeclareSIUnit{\AU}{AU}
\DeclareSIUnit{\h}{h}

\begin{document}
\thispagestyle{empty}
\raggedright
\huge
Astro2020 APC White Paper \linebreak

WAET: low-cost ground based telescopes for accelerated exoplanet direct imaging
\linebreak
\normalsize

\noindent \textbf{Thematic Areas:} \hspace*{60pt} \text{\rlap{\xmark}}$\square$ Ground Based Project \hspace*{10pt} $\square$ Space Based Project \hspace*{20pt}\linebreak
$\square$ Infrastructure Activity \hspace*{31pt} $\square$ Technological Development Activity \linebreak
  $\square$ State of the Profession Consideration \hspace*{65pt} \linebreak
  
\textbf{Principal Author:}

Name: Benjamin Monreal
 \linebreak						
Institution: Case Western Reserve University
 \linebreak
Email: \texttt{benjamin.monreal@case.edu}
 \linebreak
Phone:  216-368-0222
 \linebreak
 
\textbf{Co-Authors:} \linebreak
Dominic Oddo, Case Western Reserve University\linebreak
Christian Rodriguez, UC Irvine
\linebreak

\textbf{Co-Signers:} \linebreak
Reed Riddle, Caltech\linebreak
Chuanfei Dong, Princeton
\linebreak

\textbf{Whitepaper Description:}\linebreak
The Wide Aperture Exoplanet Telescope (WAET) is a new ground-based optical telescope layout with an extremely asymmetric aperture, which results in new exoplanet imaging reach at very low cost. We suggest that hWAET, a \SI{100x2}{\m} telescope, can be built for \$150M in the 2020s, and \SI{\ge 300}{\m} versions merit further R\&D.

\clearpage 
\setcounter{page}{1}
\raggedright
\huge
WAET: low-cost ground based telescopes for accelerated exoplanet direct imaging
\linebreak

\normalsize
\justifying

\section{Key science goals and objectives}
The Wide Aperture Exoplanet Telescope (WAET) is a recently-proposed\cite{2018JATIS...4b4001M} ground-based optical telescope layout in which {\em one} dimension of a filled aperture can be made very, very large (beyond \SI{100}{\m}) at low cost and complexity.   With an unusual beam path but otherwise-conventional optics, we obtain a fully-steerable telescope on a low-rise mount with a fixed gravity vector on key components.   Numerous design considerations and scaling laws suggest that WAETs can be far less expensive than other giant segmented mirror telescopes.  In particular, we suggest that WAET telescopes have simple enough R\&D needs, and low enough costs, that a telescope with \SI{100}{\m}-class resolution can be designed, funded, and constructed in the 2020s, and that early studies of much-larger-scales (we suggest \SI{300}{\m}) are reasonably likely to bear fruit.

The 2000 Decadal Survey\cite{astro2000}, in first recommending construction of \SI{30}{\m} telescopes, expressed hope that they'd cut costs (relative to an extrapolation from Keck with a $D^{2.6}$ cost/diameter scaling law) by a factor of 4.  With regards to the then-active \SI{100}{\m} OWL project, the survey said the following:
\begin{quotation}
In comparison, to build the much more powerful ESO 100-m OWL ... for \$1.5 billion will require innovation that reduces costs relative to Keck by a factor of 20. Such a cost reduction is significantly more challenging and does not appear to be reasonable for a single engineering step. In fact, if OWL could be built for this price, the same technology could produce a 30-m telescope for \$65 million, which would be an even more compelling next step. The panel is excited about the possibility of OWL but expects that it will probably take much more time to be developed than GSMT.
\end{quotation}

We believe that WAET \emph{is} a reasonable single step to post-30-m-class science, at reasonable costs and using existing technology.  

Astro2020 science whitepapers reflect the community interest in reflected-light images of rocky exoplanets\cite{2019BAAS...51c.128M,2019BAAS...51c.200W,2019BAAS...51c.514M}, and a general sense that extreme AO coronagraphy is up to the challenge whenever the planet-star separation is adequate.  In this whitepaper, we argue that a \SI{100}{\m}-class WAET facility we call ``hWAET'' can reach high enough resolutions to deliver many rocky planet images this decade, at a reasonable cost, with low R\&D risk.  Although WAET conceptual design, science-case development, and costing are at a very early stage, we have identified no showstoppers and believe the project merits community consideration.

\section{Technical overview}
 
In this section, we review the WAET mount.  WAET telescopes have a highly elongated pupil, with one long dimension $L$ and one short dimension $W$ for an aperture of $A = L\times W$.  The key cost-saving measure is that WAET can be steered without elevating the long axis of any component.   A flat slit-shaped siderostat provides all altitude steering, while the slit-shaped primary translates along the ground without tilting to steer in azimuth.    We have done mechanical and optical design exercises for one implementation (hectometer-WAET or ``hWAET'' at \SI{100 x 2}{\m}) which we argue is a low-risk route towards ground-based direct imaging of rocky exoplanets; and for a much larger instrument (``kWAET'' at \SI{300 x 5}{\m}) with post-TMT-class light collection and sub-milliarcsecond resolution, which would be affordable with current technology but which is large enough to benefit from OWL/HET/SALT-like low cost mirror R\&D.   The asymmetric pupil leads to a highly elongated PSF, which appears tolerable for general-purpose astronomy and which is beneficial for exoplanet searches since it reasonably often aligns with the planet-star separation.

\begin{figure}
\begin{center}
\includegraphics[width=0.7\textwidth]{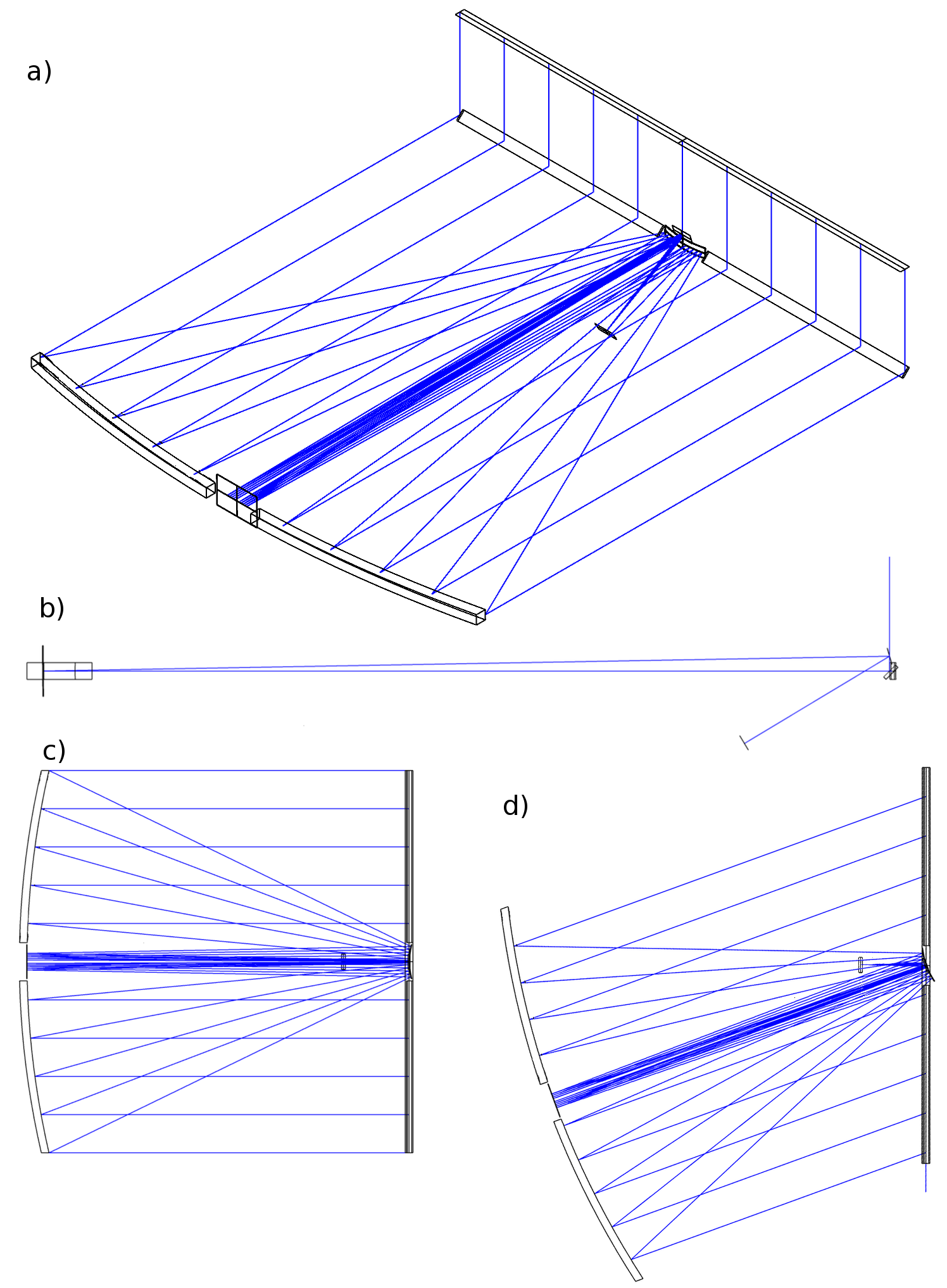}
\end{center}
\caption{WAET optical layout.   Starlight (shown as coming from zenith) reflects once off a tilted siderostat, then off a focusing primary.   The details shown are of an f/1.1 Ritchey-Chr\'etien layout, with the secondary mounted in a gap in the siderostat and a flat tertiary fixed to the primary.  A pickoff mirror above the secondary sends light to instruments below the beam plane.   a) isometric view. b) side view.   c) top view with the telescope viewing a source at zenith.  d) top view illustrating the ``slew'' positioning of the primary and secondary; the telescope is viewing a target 20$^\circ$ north of zenith.}\label{fig1}
\end{figure}

We will survey the basic WAET operating principles and preview some of the known advantages and disadvantages of the design.  Fig.~\ref{fig1} shows an optical model, specifically of a Ritchey-Chr\'etien implementation.

Starlight is incident on a flat siderostat (M1) whose only degree of freedom is \emph{roll} around the long axis.  The siderostat folds the beam into the horizontal plane and directs it towards the horizon-facing primary (M2).   The primary focusing mirror is, like the siderostat, a thin elongated shape with its short axis vertical and its optical axis parallel to the ground. The primary mirror moves with one degree of freedom: sliding horizontally on a bearing, it executes \emph{slew} about a vertical axis near the center of the siderostat.    Siderostat-roll and primary-slew, working together, steer the telescope's optical axis freely across most of the sky.  No structures ever elevate far from the ground plane.  

WAET can be seen as a fully-steerable optical variant of the Kraus-type radio telescope, notably implemented as the Big Ear at Ohio State (1963--1998)\cite{Kraus:SkyAndTelescope:1953} and the Nan\c{c}ay Radio Telescope \mbox{(1965--)\cite{Lequeux:JournalOfAstronomicalHistoryAndHeritage:2010}}.   In contrast to WAET, Kraus-type telescopes have a non-tracking siderostat; they operate at fixed elevation, either as transit telescopes or with a moveable (\SI{15}{\degree\per\hour}) feed that can track targets briefly at the chosen elevation.  With a slewable primary, WAET obtains conventional tracking and substantial sky coverage (see section \ref{label_skycoverage}).   (Similar performance is obtained with a stationary primary and an alt-az mounted siderostat, among other variants.)

Our system is compatible with various well-understood optical prescriptions. In this whitepaper, most diagrams show a rectangular-aperture Richey-Chr\'etien, but a prime focus Newtonian telescope might be optimal for a more narrowly exoplanet-optimized facility which tolerates a small field of view.  In all designs, we can easily specify a stationary focal plane at ground level, allowing instruments to occupy nearly-unlimited space and weight.

\subsection{Mechanical design for low cost} The WAET layout allows us to use extremely simple, lightweight mechanical structures.   The primary mirror is mounted on a low-rise, non-tilting structure, like a long curved wall; it has a constant gravity vector and does not flex (except, at few-micron level requiring active figure control, due to bearing non-flatness) during tracking, when it ``slews'' by sliding along a long curved foundation pad.  The siderostat, although it tilts while tracking, requires mechanical stiffening along a \emph{short} axis only.  Both the primary and siderostat mounts are made of identical repeated subunits, allowing economies of scale in design, manufacturing, and shipping.  The telescope needs no standard dome; an open design is possible, requiring only shedlike protection of the long structures, or low-thermal-mass membranes may be stretched above and/or below the beam path.  The bearing pads, which represent most of the civil engineering, are more nearly comparable to low-cost warehouse flooring than to a standard giant-telescope pier.

We have attempted to estimate WAET's costs and cost scaling laws; despite many uncertainties, it appears likely that WAET's total construction cost is dominated by the cost of figured mirrors and mirror cells.  Since segmented-mirror production is a mature technology, and WAET's primary segments do not demand any novel or cost-uncertain engineering (indeed, most specifications are looser than those of TMT/ELT segments) we are able to estimate WAET project budgets with some confidence even at this early stage.  Less-predictable elements of the project cost (including mount manufacturing, civil engineering, and siderostat flat mirror production) appear to be small corrections on top of the predictable primary-mirror cost. 

\subsection{PSF and AO performance}
From an observer's perspective, WAET's most unusual feature is its asymmetric PSF, which is elongated in one direction by factors of 10.  The reader may judge how annoying this is for their own observations.  Since both telescope axes are larger than a Fried length, the shape does not seem to have qualitatively-important adaptive optics implications, except possibly the loss of frozen-flow predictive control in AO loops.  Other than the central obstruction, the beam path is perfectly clear, with no secondary-mirror support spider, which may aid coronography\cite{macintosh_extreme_2006}.  WAET has an unusual full-telescope autocollimation mode between the primary and siderostat; we hope this will permit unusually-precise figure control and suppression of static speckles.  

Unfortunately, WAET's light path traverses a considerable distance near the ground---at least twice the focal length.  The ``dome seeing'' along this path might be a serious issue, and it is almost certainly the case that WAET has worse natural seeing than conventional telescope in a modern dome on an elevated pier.   Discovering \emph{how much worse} it is, and what interventions are possible (Beam path elevation?  Enclosure?  Siting?  Active sensing?), is a critical study topic both for the science case and the engineering design.  We discuss some rough AO estimates in section~\ref{section_exo}.

\subsection{Sky coverage}\label{label_skycoverage}

\begin{figure}\begin{center}
    \includegraphics[width=0.7\textwidth]{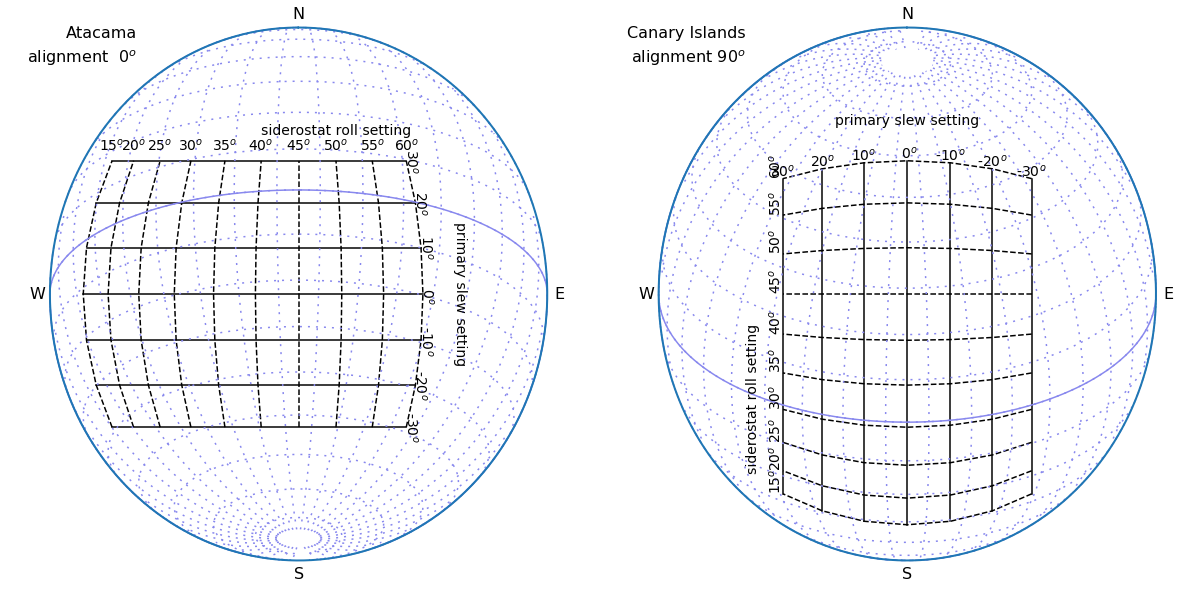}
\end{center}
\caption{Sky coverage and pointing parameters for two illustrative WAET site choices.  In both cases, the area shown is unvignetted assuming a siderostat $2\times$ taller and $1.15\times$ wider than the primary.}\label{skycov}
\end{figure}

WAET sky coverage includes a wide ``stripe'' of sky, generally including zenith.  Behind the primary, the sky is visible at low elevation; rolling the siderostat steers the beam along the stripe towards zenith or beyond. A siderostat $\sqrt{2}\times$ taller than the primary can reach zenith unvignetted; a siderostat $2\times$ wider than the primary can reach 30$^\circ$ past zenith.  The width of the stripe is determined by the primary mirror's slew range, which might be limited by mechanical stops or by vignetting of the siderostat.  A siderostat $1.15\times$ longer than the primary allows primary slew to $\pm 30^\circ$ without vignetting.  Two example skies are shown in Fig. \ref{skycov}.  Note that an E-W facing telescope steers with a fast roll and a slow slew, while a N-S facing telescope requires a fast slew and slow roll.  Field rotation is usually small.  \cite{2018JATIS...4b4001M} described some WAET variants with different coverages.

\subsection{Specific implementations: hWAET and kWAET}

\subsubsection{hWAET (hectometer-WAET)}

\begin{figure}
\begin{center}
\includegraphics[width=0.75\textwidth]{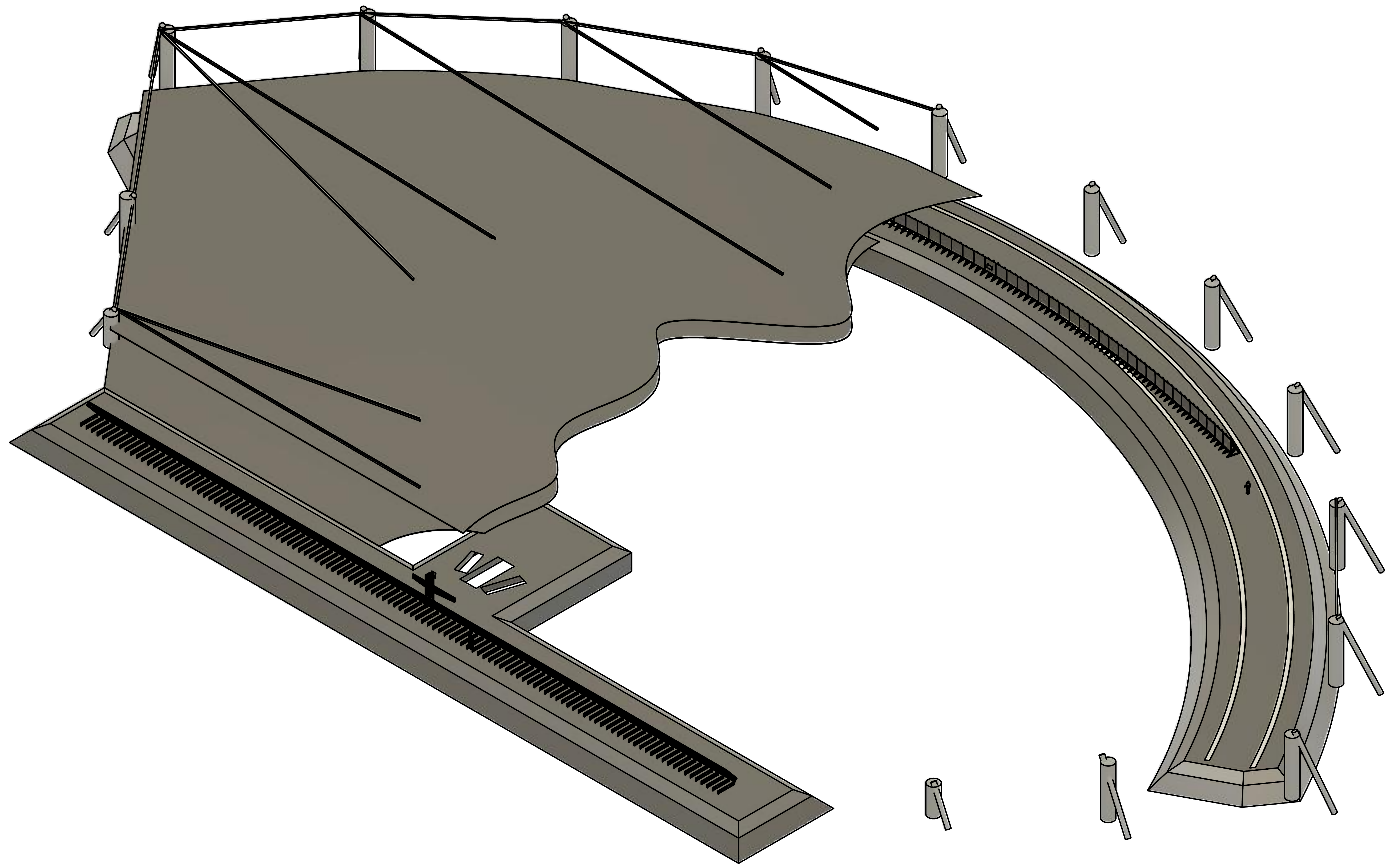}
\caption{Isometric sketch of a hWAET installation with a tightly-enclosed beam.  At bottom left is the \SI{115 x 3}{\m} siderostat; at the upper right is the \SI{100 x 2}{\m} primary mirror on its long bearing platform.  The Richtey-Chr\'etien secondary is a wide rectangle near the siderostat center.  Note a \SI{1.8}{\m} person for scale (near right end of primary).  Low-thermal-mass windproof membranes are stretched above and below the beam path, suppressing advected turbulence.}\label{fig_hwaet_mech}
\end{center}
\end{figure}

A \SI{100 x 2}{\m} aperture (hectometer-WAET or hWAET) (fig. \ref{fig_hwaet_mech}) is the system which, we argue, has an attractive budget, scope, and timeline for the 2020s.  It takes advantage of the WAET layout and realizes post-\SI{30}{\m} science capabilities, but otherwise has low R\&D risks.  The \SI{2}{\m} dimension allows complete primary and siderostat subassemblies to fit in shipping containers.  Fig. \ref{fig_hwaet_mech} shows hWAET in a Richey-Chr\'etien configuration with an f/1.1 primary, f/27 secondary, and instrument rooms below the beam plane.  hWAET's \SI{200}{\m\squared} collecting area is equivalent to a \SI{16}{\m} circular aperture and it has a  \SI{2}{\mas} diffraction limit at \SI{1}{\micro\m} (the separation of, e.g., TRAPPIST-1$e$).  

\subsubsection{kWAET (kilometer-WAET)}

\begin{figure}
\begin{center}
\includegraphics[width=0.85\textwidth]{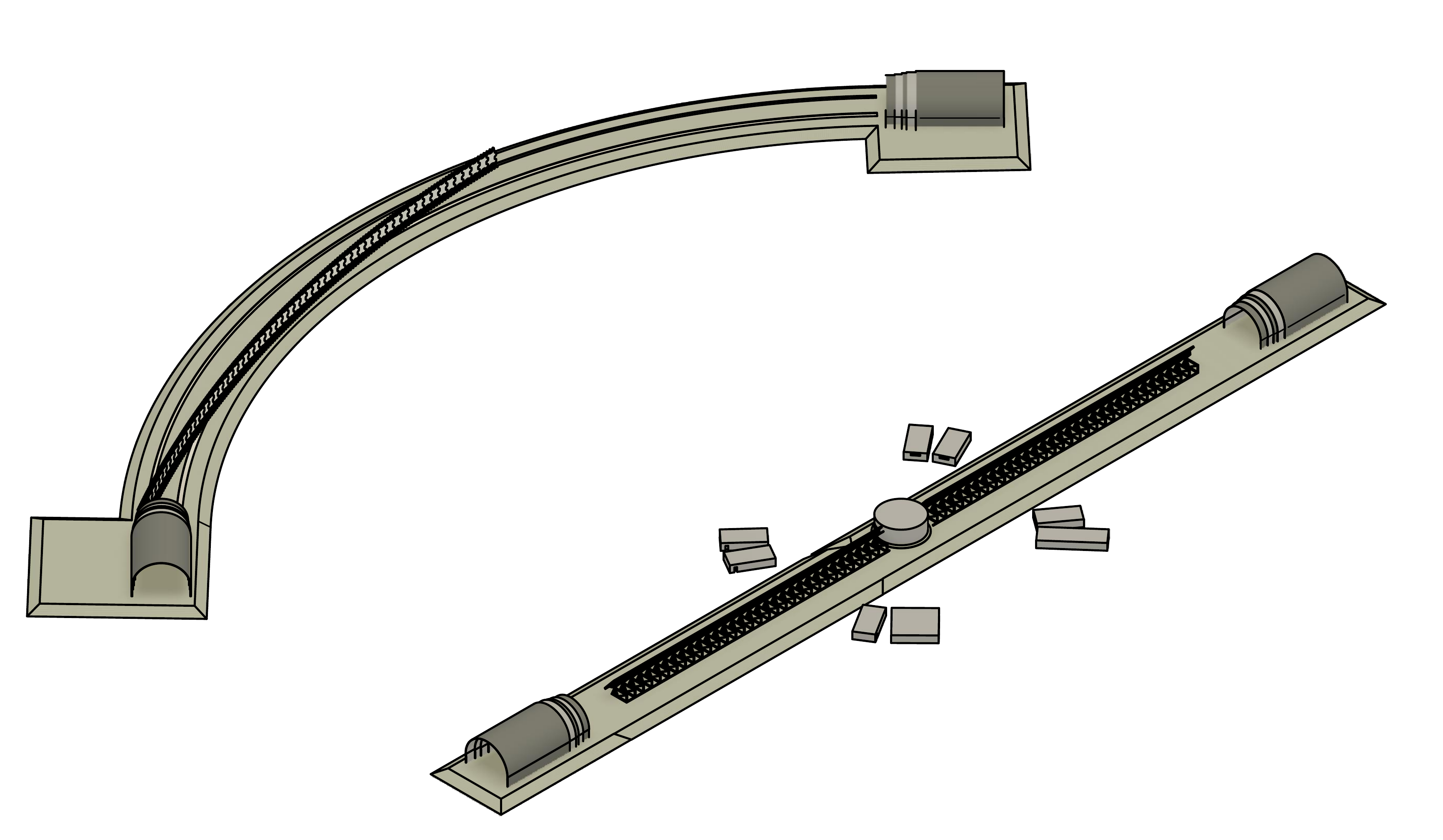}
\caption{Isometric sketch of an open-air kWAET installation.  At the top is the curved bearing holding the \SI{300 x 5}{\m} primary; the primary is shown at 15$^\circ$ slew to the left.   At the bottom is the linear pad holding the \SI{345 x 8}{\m} siderostat, with a secondary package in a gap.  Nested half-round sheds extend into enclosures.  For scale, note \SI{2}{\m} doors in the camera buildings.}\label{fig_kwaet}
\end{center}
\end{figure}

There is no obviously insurmountable barrier to a WAET telescope approaching kilometer scale (kWAET).   Subject to further optimization, consider a \SI{300 x 5}{\m} aperture (Fig.~\ref{fig_kwaet}).  This has the collecting area of a \SI{50}{\m} telescope (or 3x TMT) and a \SI{0.7}{\mas} diffraction limit (below, e.g., TRAPPIST-1b).  The primary could be 477 \SI{2.2}{\m} hexagonal segments in a \num{3 x 159} grid, mounted on 53 identical nine-mirror subassemblies.  The siderostat segments are 138 \SI{3.4 x 5}{\m} rectangles, installed on 69 identical \SI{6.8 x 5}{\m} subassemblies.   A tensioned roof at this scale is difficult, so kWAET would either be an open-dome design or would have to accept thin roof supports in the beampath.  Without extensive grading, kWAET could be sited flat at, e.g., Llano de Chajnantor, slightly tilted at Magdalena Ridge, or steeply angled (losing the ``constant gravity vector'' claims) on, e.g., the northwest face of Cerro Paranal.

\section{Science capabilities}

Understanding the science case for a billion-dollar telescope, quite reasonably, involves years of work by hundreds of people.  In this document, we can only preliminarily sketch a very early picture of the possible science case for WAET.   Our simulation effort to date has focused on exoplanets, which have a clear and unique connection to WAET's aperture shape.   Many other topic areas, particularly spectroscopic ones, are involved insofar as WAET would implement very large collecting areas with an AO-limited PSF; for many targets WAET can be an all-purpose workhorse GSMT at whatever total area is requested.  

\subsection{Exoplanet imaging}\label{section_exo}

We now know that planetary systems are a common feature of stars in the Milky Way.  While radial-velocity and transit surveys have made it possible to do \emph{populations studies} of planets as a class of objects, it is of the utmost scientific and popular interest to study a few of them in as much detail as possible, including in reflected light.  WFIRST and \SI{30}{\m}-class telescopes should be able to do this routinely with exo-Jupiters but only rarely (and always at the edge of resolvability) capture super-Earths or Earths.  Making some projections about the extreme-AO limitations of WAET's aperture shape, we conclude that it has excellent rocky exoplanet imaging reach.

\begin{figure}[h!]
\begin{center}
\begin{minipage}{0.75\linewidth}
\includegraphics[width=\textwidth]{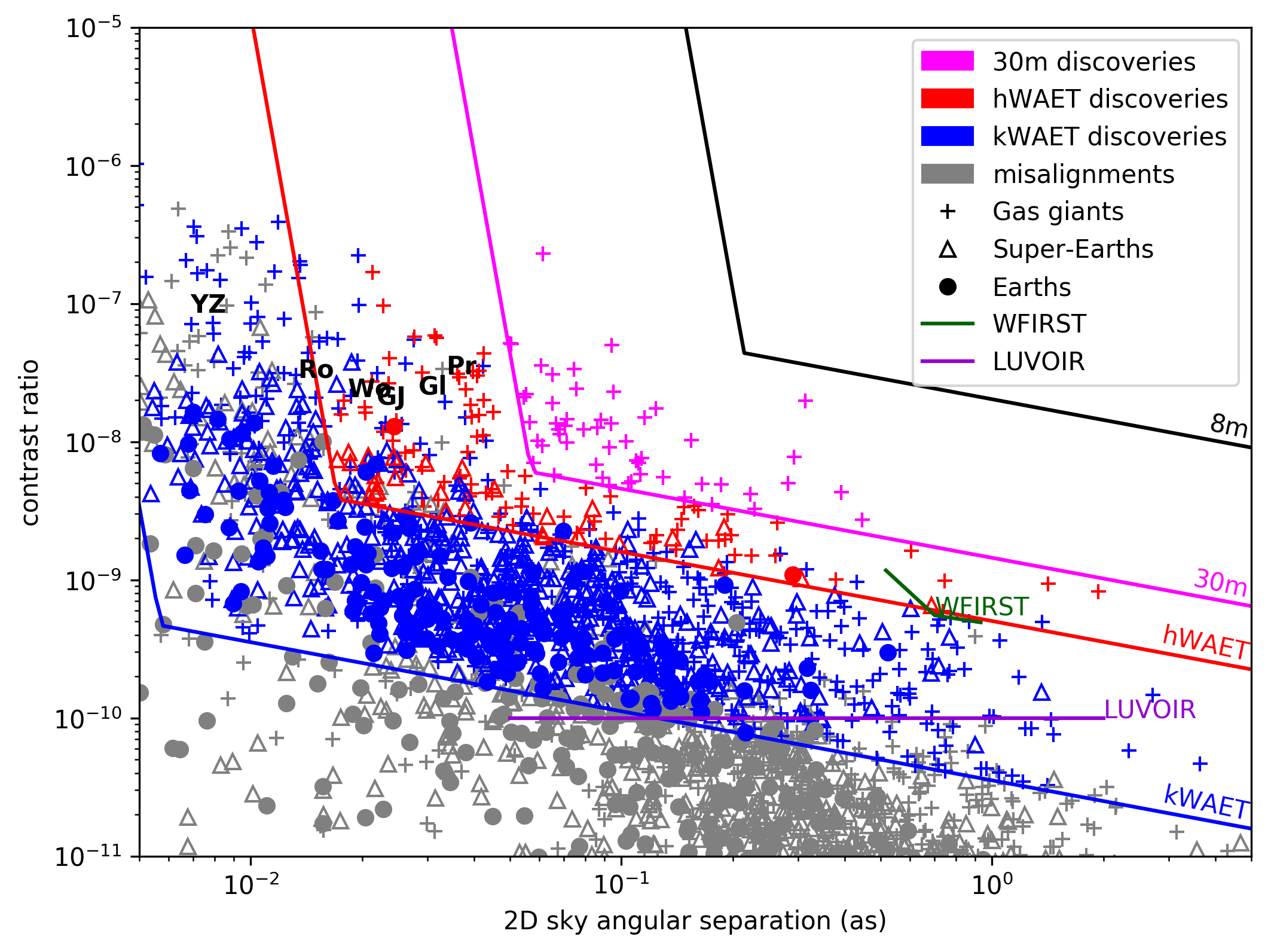}
\end{minipage}
\begin{minipage}{0.2\linewidth}
\includegraphics[width=\textwidth]{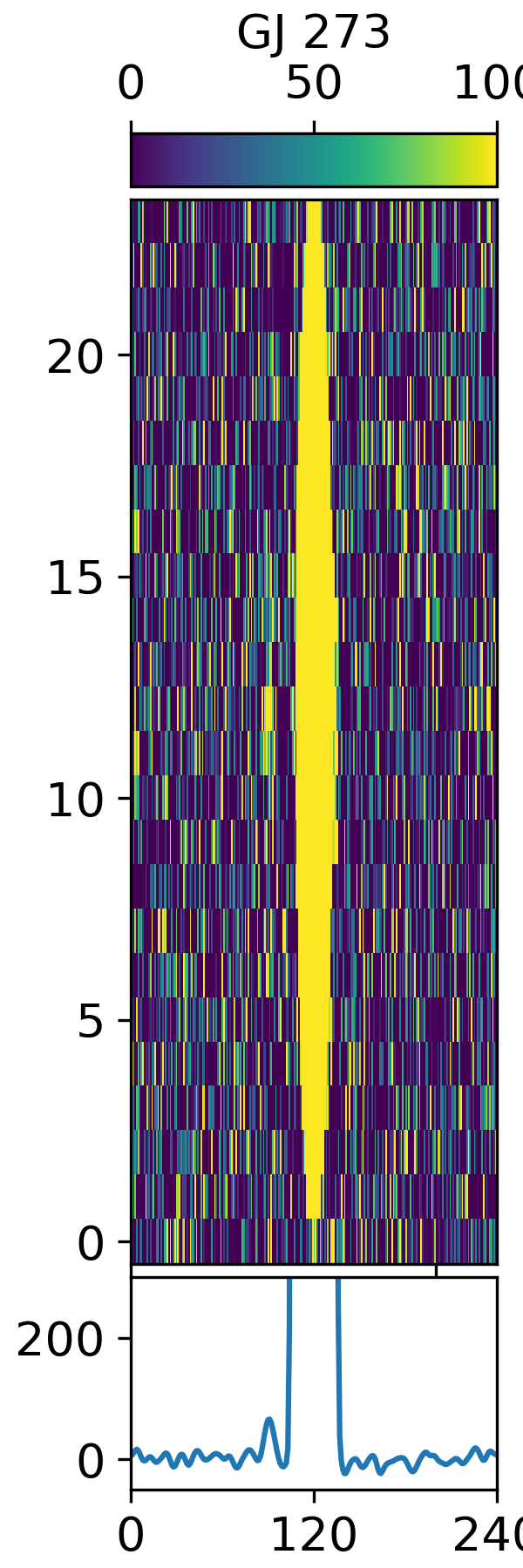}
\end{minipage}

\caption{Left: A preliminary, unoptimized sky survey showing hWAET and kWAET discovery potential.  From the EXOCAT-1 catalog of stars within 30~pc, we select 900 with $-55 < dec < 5$ visible from Paranal with a typical WAET-like slew limit (see Fig \ref{skycov}, left).  The x-axis shows the actual angular separation on the sky for the (random) observation epoch, but detectability is calculated using a projection of this onto WAET's one high-resolution axis, resulting in some missed detections (shown in grey) which might or might not become detectable in a different orbital phase.  The targets suggested by \cite{2019BAAS...51c.128M} are labeled.  Right: \SI{100}{h} I-band hWAET image of GJ 273 with $c$ clearly resolved on the left.}\label{exosims}\end{center}
\end{figure}

To estimate WAET's exoplanet discovery reach, we first need to predict its xAO performance limits.  To obtain very preliminary estimates, we use the methods of \cite{guyon_limits_2005} to calculate the ``raw'' PSF contrast, for \SI{100}{\m} and \SI{300}{\m} circular apertures, assuming an ideal ($\beta==1$) wavefront sensor, and we degrade by a factor of a few to account for both ground-level seeing and the absorption of additional spatial modes into WAET's asymmetric PSF.  Starting with the EXOCAT-1 catalog and a Kepler-2 simulated planet population in EXOSIMS\cite{delacroix_science_2016}, we label as ``discoveries'' as planets whose contrast is 2.5x greater than the PSF contrast, and at conservative separation $\alpha > 5 \lambda/D$ (projected on the high-resolution axis) at $\lambda$=\SI{1.25}{\micro\m}; in this we mimic the treatment used in, e.g., \cite{national_academies_of_sciences_exoplanet_2018}, recognizing that more-optimistic AO practitioners discuss far smaller IWAs.   Our results are shown in Figure \ref{exosims} and compared with \SI{30}{\m} xAO, LUVOIR, and WFIRST sensitivities.  With no survey optimization, we conclude that hWAET would image tens of rocky planets.  Note that multiple close-orbit Earth-mass objects become detectable even if xAO progress were stalled along the \SI{30}{\m} projection; Prox Cen b, Gliese 411, GJ 273b, Wolf 1061 c, and Ross 128 b are accessible up to the luck of alignment.  Further AO contrast improvements will help to extend the discovery space towards the habitable zones around hotter stars.  kWAET would have access to $\sim$ 150 Earth-mass and $\sim$ 400 super-Earth-mass planets.  While this paragraph has focused on rocky planets, a large sample of reflected-light gas giants is expected.   We can conclude that hWAET would make important exoplanet discoveries in the late 2030s, even in a world with WFIRST and multiple \SI{30}{\m} telescopes in operation.

Further study of WAET exoplanet spectroscopy and characterization capabilities is in progress.  Here we note that hWAET's asserted \SI{2}{\m} height, chosen fairly arbitrarily, is easily raised if the collecting area is inadequate or lowered for budgetary reasons.

\subsection{WAET as a general-purpose telescope} 

WAET's single long axis provides special benefits in science areas which, like exoplanets, have some important feature laid out linearly on the sky.  Beyond exoplanets, we might consider for particularly interesting applications to, e.g., high-z galaxy kinematics\cite{glazebrook_dawes_2013}; Galactic Center astrometry; imaging of AGN tori; the authors welcome other applications.  Targets with notable 2D morphology might suffer.

When studying point sources, WAET's unusual shape does not matter; its PSF covers a very small solid angle, inversely proportional to the aperture area, which yield the standard giant-telescope properties of low confusion in crowded fields and low sky noise.  Unlike a circular telescope of the same area, WAET can sometimes exploit field rotation for a sort of ``tomographic'' reconstruction of crowded fields or fine 2D detail.   Therefore, insofar as the unusual PSF and sky coverage are tolerable, WAET is worth building for general-purpose visible, IR, and MIR imaging and spectroscopy.   In this domain, WAET should be compared to conventional alt-az telescopes on the basis of area per cost, where we project a fairly linear \SI{1300}{\m\squared}/\$1B (assuming off-the-shelf GSMT mirror technology and costs) or \SI{5200}{\m\squared}/\$1B (if the OWL/SALT/HET-like option is viable).   This is to be compared with something around \SI{500}{\m\squared}/\$1B for today's \SI{30}{\m} projects.


\section{Specific project suggestions and cost estimation}

Cost figures (2017 USD) come from our published cost estimation exercise\cite{2018JATIS...4b4001M}, which excludes instruments.  The author notes that the extreme scarcity of publicly-accessible \emph{project management data and budgets} made this exercise unnecessarily difficult and uncertain.  

\paragraph{1: Clarify the exoplanet science case for 100~m and beyond} The first recommendation for the Astro2020 community is for a multi-PI design and performance study which will clarify the exoplanet science case for post-\SI{30}{\m} giant apertures, of which WAET may be only one example.

\paragraph{2: Design, propose, build hWAET for fast track direct imaging in the 2020s}  If preparatory work, with science-based optimized specifications, begins in 2020, we believe a hWAET-scale telescope project can be launched (tapping recently-idled GMT/ELT/TMT design expertise), designed, funded, and built before 2030, although a detailed timeline is unavailable. Our cost model projects a hWAET construction budget around $\sim$150M, dominated by $\sim$100M for primary mirror and $\sim$17M for the siderostat, and we believe the R\&D risks are low.  Unless conceptual-design work turns up surprises, hWAET might be the fastest track towards habitable-zone planet images, and at a budget which does not crowd out other approaches.  

\paragraph{3: The kWAET scale justifies revived mirror manufacturing R\&D} To build kWAET with low R\&D risk and high cost certainty, we might simply extrapolate current-technology GSMT mirror costs, in which case we project \$1.15B, which is clearly within precedent for (rare, slow-to-develop) large international telescope projects.  However, recall that spherical-primary designs exist \cite{1998SPIE.3352..778G}\cite{ray_issues_2000}\cite{sebring_extremely_1998} and continue to promise lower-cost mirror mass production.  While spherical-mirror costs were \emph{one of many} technical and budget risks for OWL itself, we feel that kWAET illustrates an approach where nearly 100\% of the budget risk (and indeed nearly 100\% of the budget) can be attributed to primary segment procurement.  With OWL-derived costing (and attendant uncertainty) our model projects a spherical kWAET to cost \$210M--\$280M, a far easier and quicker funding proposition and one which permits some risk tolerance.   Therefore, if the 2020s see a revival of cheap spherical optics, clarifying their costs and lowering the perceived technical risk, \emph{with considerable certainty} the results translate, via kWAET, to affordable gigantic apertures and deep exo-Earth imaging in the 2030s.  We do not have a detailed cost estimate or timeline for this line of R\&D.

\begin{figure}
\begin{center}
\begin{minipage}{0.45\linewidth}
\includegraphics[width=0.85\textwidth]{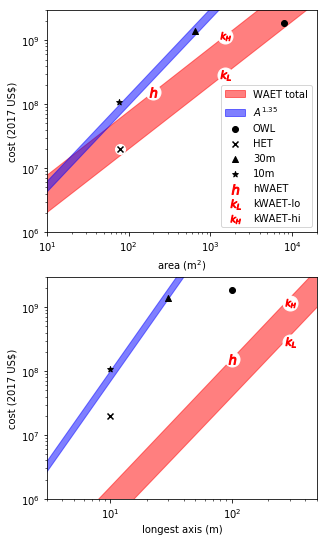} 
\end{minipage}
\begin{minipage}{0.45\linewidth}
\begin{tabular}{ l   r l } 
\hline
\multicolumn{3}{c}{\bf{Cost scaling estimates}}\\
\hline
\multicolumn{1}{c}{Component} & \multicolumn{2}{l}{Scaling estimate}\\
\hline
Superstructure          & \$3k & L$\times$W\\
Slew/roll bearings & \$2k & L$\times$W\\
Figure control   & \$10k & L$\times$W\\
Enclosures  & \$2k & L$\times$W$^{0.5}$\\
Flat mirrors       &  \$58k  & L$\times$W\\
\addlinespace
\multicolumn{3}{l}{Primary mirror:} \\    
  a) Keck-like & \$450k  & L$\times$W\\
  b) OWL-like & \$50k &  L$\times$W\\
\addlinespace
\multicolumn{3}{l}{Slabs and bearing:} \\
  a) standard     & \$250 & L$^{1.6}$\\
  b) alt-az siderostat & \$150 & L$^{2}$ \\
\addlinespace
\multicolumn{3}{l}{Beampath interventions:} \\
  a) Ground insulation   & \$40 & L$^2$\\
   b)                       Propped roof      & \$200 & L$^2$  \\
    c)  Stretched roof      & \$1k & L$^2$  \\
\hline
\end{tabular}
\end{minipage}
\caption{WAET cost scaling laws (2017 USD) as a function of area (top) or longest-dimension (bottom), assuming a 50:1 aspect ratio.  The red band reflects the range between low-cost and standard mirrors.  For reference, we show ballpark totals for: HET; typical 8m and 30m telescopes; late OWL projections; hWAET; and low- and high-cost versions of kWAET.  The blue band shows a typical $A^{1.35}$ scaling law.  Scaling-law breakdowns (right) show that mirror costs dominate despite large component uncertainties.}\label{fig_cost}\end{center}
\end{figure}
\clearpage
\printbibliography
\end{document}